\documentstyle[aps,pra,twocolumn,epsfig,floats]{revtex}

\def\Red  {}
\def\Black{}
\def\Blue {}
\begin{document}
\draft
\title{\Red Formation of ultrashort pulses with sub-Poissonian\\ photon
statistics\Black
 }
\author{F. Popescu${}^{*}$\and A. S. Chirkin${}^{\dagger}$}
\address{Moscow State University,\\
         119899 Moscow, Russian Federation\\  [1mm]
        {\sf E-mail: ${}^{*}$florentin\_p@hotmail.com,
                ${}^{\dagger}$chirkin@foton.ilc.msu.su}}
\date{March 10, 2000}
\twocolumn[
\widetext
\begin{@twocolumnfalse}
\maketitle
\begin{abstract}\Blue
A simple method for the production of ultrashort light pulses (USPs) with suppressed
photon fluctuations is considered. The method is based on self-phase modulation (SPM)
of an USP in a nonlinear medium  (optical fibre) and subsequent transmission of
pulse through a dispersive optical element.\Black\\
\medskip
{\sf Enlarged version of the article published  in  \emph{Quantum Electronics}, {\bf 29} No. 7, 61-63 (1999)}
\end{abstract}
\vspace{0.5cm}
\narrowtext
\end{@twocolumnfalse}]
Ultrashort light pulses (USPs) continue to draw the attention of investigators today.
The state of the art in this area of laser physics in the late 1980s was set forth by
S. A. Akhmanov \emph{et al}. in Refs.\cite{Akhmanov},\cite{Vysloukh}. In the past decade,
considerable progress has been made in quantum optics in the generation of
nonclassical (the so called squeezed) light fields. The formation and application of
USPs in a nonclassical state makes possible to combine in experiments a high time
resolution with a low level of fluctuations.

In principle, one can obtain pulsed light fields in a nonclassical state by using the
same nonlinear optical interaction as those used in the case of continuous fields
\cite{Belinski1},\cite{Wals}. Parametric amplification is a technique that is most
extensively used for this purpose nowadays. In the case of degenerate three-frequency
parametric amplification, quadrature-squeezed light is produced. However, this light
is found to have super-Poissonian statistics directly at the output of the amplifier,
and one needs interferometers to transform it to get sub-Poissonian statistics. One
can obtain light with sub-Poissonian photon statistics with the aid of nonlinear
interferometric devices in the presence of self-phase modulation (see, e.g.,
Ref.\cite{Belinski1}). Note that the self-phase modulation itself is not accompanied
by a change in photon statistics.

 In a recent paper, we studied SPM of a light pulse and its subsequent propagation in
a dispersive linear medium (or the passage through optical compressors). In this
case, pulses with sub-Poissonian photon statistics can be formed. We  managed to make
an accurate calculation of the process under consideration owing to the consistent
quantum theory developed by us for the self-action of light pulses in a medium with
inertial nonlinearity \cite{Popescu1},\cite{Popescu2}. The method of formation of pulses
with sub-Poissonian statistics pro\-po\-sed here is simple for experimental realization
and stable against external random effects (technical fluctuations). Note
that the theory developed in Refs.\cite{Popescu1},\cite{Popescu2} for the self-effect
of USPs takes into account a relaxation time of a nonlinearity that determines the
region of the spectrum of quantum fluctuations which are of substantial importance in
the formation of squeezed light and does not limit the amplitude of quantum
fluctuation.

 When analysed from the quantum point of view, SPM of USPs is described by the
expression \cite{Popescu1},\cite{Popescu2}.
\begin{equation}\label{oo}
\hat{A}(t,l)=e^{\hat{O}(t)}\hat{A}_{0}(t),
\end{equation}
and the Hermite conjugate operator $\hat{A}^{+}(t,l)$. Here $\hat{A}^{+}(t,l)$
($\hat{A}(t,l)$) is the photon creation (annihilation) operator for the given section
$x$ at a given moment of time $t$ (the output of a nonlinear medium is specified by
the section $x=l$, and the input is specified by the section $x=0$,
$\hat{A}(t,x\smash{=}0)=\hat{A}_{0}(t)$), $\hat{O}(t)=i\gamma q[\hat{n}_{0}(t)]$, and
$\gamma$ is the coefficient related to nonlinear properties of a medium and
proportional to the length $l$ in it. The expression $\gamma q[\hat{n}_{0}(t)]$
characterizes the nonlinear phase incursion,
\begin{equation}\label{qu}
q[\hat{n}_{0}(t)]=\int\limits_{-\infty}^{\infty}H(|t_{1}|)\hat{n}_{0}(t-t_1)dt_1,
\end{equation}
and $\hat{n}_{0}(t)=\hat{A}^{+}_{0}(t)\hat{A}_{0}(t)$ is the operator of the
"density" of the number of photons at the input of a nonlinear medium. The function
$H(t)$ takes into account a finite response time of the nonlinearity of a medium. In
view of the causality principle, $H(t)\neq0$ for $t\geq0$ and $H(t)=0$ for $t<0$. In
expression (\ref{oo}) and (\ref{qu}), $t=t^{'}-x/u$ is the time in the moving
coordinate system, $t^{'}$ is the current time, and $u$ is the pulse velocity in a
medium.

In the case of SPM (see (\ref{oo})) discussed here, the quadrature-squeezed light
\cite{Popescu1},\cite{Popescu2} is formed, and the photon statistics in a medium is
not changed because the operator of the number of photons
$\hat{n}(t,l)=\hat{n}_{0}(t)$ remains unchanged. In what follows, we show that the
propagation of a pulse (see (\ref{oo})) through a dispersive optical device is able
to transform its initial Poissonian statistics into sub- or super-Poissonian
statistics.

For example, let us consider the propagation of a pulse in a dispersive linear
medium. At the output of the medium, we have the following expression for the photon
annihilation operator \cite{Vysloukh}:
\begin{equation}\label{be}
\hat{B}(t,z)=\int\limits_{-\infty}^{\infty}G(t-t_{1},z)\hat{A}(t_{1},l)dt_{1}.
\end{equation}
It is known that the operators $\hat{A}(t,l)$ and $\hat{B}(t,z)$ are bound to satisfy
commutation relation of the $[\hat{C}(t_{1}),\hat{C}^{+}(t_{2})]=\delta(t_{1}-t_{2})$
type. In view of relation (\ref{be}) between the operators under consideration, we
find the condition imposed on the Green function $G(t)$ of the dispersive element:
\begin{equation}\label{delta}
\int\limits_{-\infty}^{\infty}G(t_{1}-t,z)G^{*}(t_{2}-t,z)dt=\delta(t_{2}-t_{1}).
\end{equation}
Let us introduce the operator of the number of photons over the measurement time
${\mathcal{T}}$,
\begin{equation}\label{en}
\hat{N}_{{\mathcal{T}}}(t,z)=\int\limits_{t-{\mathcal{T}}/2}^{t+{\mathcal{T}}/2}
\hat{N}(t_{1},z)dt_{1},
\end{equation}
and define the Mandel parameter
\begin{equation}\label{par}
Q(t,z)=\frac{\varepsilon(t,z)}{\langle\hat{N}_{{\mathcal{T}}}(t,z)\rangle}\, ,
\end{equation}
where
\begin{eqnarray}
\hat{N}(t,z)&=&\hat{B}^{+}(t,z)\hat{B}(t,z);\nonumber\\
\varepsilon(t,z)&=&\langle\hat{N}^{2}_{{\mathcal{T}}}(t,z)\rangle-
\langle\hat{N}_{{\mathcal{T}}}(t,z)\rangle^{2}-
\langle\hat{N}_{{\mathcal{T}}}(t,z)\rangle.\label{bra}
\end{eqnarray}
The parameter $Q(t,z)$ characterizes the difference between photon statistics and
Poissonian statistics (for the latter $Q(t,z)=0$). The angle brackets in (\ref{bra})
denote averaging over the initial quantum state of the pulse.

One can calculate expression (\ref{bra}) by using the algebra of time dependent Bose
operators \cite{Popescu1},\cite{Popescu2}. In this case we have the following
relations:
\begin{equation}\label{algebra}
\hat{A}_{0}(t_{1})e^{\hat{O}(t_{2})}=e^{\hat{O}(t_{2})+{\mathcal{H}}(t_{2}-t_{1})}\hat{A}_{0}(t_{1}),
\end{equation}
\begin{equation}\label{teo}
e^{\hat{O}(t)}=\hat{\mathbf{N}}\exp{\biggl\{\int\limits_{-\infty}^{\infty}
\left[e^{{\mathcal{H}}(t_{1})}-1\right]\hat{n}_{0}(t-t_{1})dt_{1}\biggl\}},
\end{equation}
where ${\mathcal{H}}(t)=i\gamma h(t)$, $h(t)=H(|t|)$ and $\hat{\mathbf{N}}$ is the
operator of normal ordering.

In the case of a coherent initial pulse, a small nonlinear phase incursion per
photon, and the pulse duration $\tau_{p}$ is considerably grater than the times
${\mathcal{T}}$ and $\tau_{r}$ ($\tau_{r}$ is the nonlinearity relaxation
time).
After some cumbersome calculations, taking into account expression (\ref{algebra}) and
(\ref{teo}) one gets
\begin{equation}\label{epsilon}
\varepsilon(t,z)=2\gamma{\mathcal{T}}^{2}Im[I^{2}_{1}(t,z)I^{*}_{2}(t,z)],
\end{equation}
where
\begin{eqnarray}
I_{1}(t,z)&=&\int\limits_{-\infty}^{\infty}G(t-t_{1},z)\alpha(t_{1})e^{i\psi(t_{1})}dt_{1};\\
I_{2}(t,z)&=&\int\limits_{-\infty}^{\infty}\int\limits_{-\infty}^{\infty}G(t-t_{1},z)G(t-t_{2},z)\alpha(t_{1})\alpha(t_{2})\nonumber\\
&\times&h(t_{1}-t_{2})e^{i\,[\psi(t_{1})+\psi(t_{2})]}dt_{1}dt_{2};
\end{eqnarray}
Here $\alpha(t)$ is the eigenvalue of the operator $\hat{A}_{0}(t)$;
$\psi(t)=2\gamma\bar{n}_{0}(t)$ is the nonlinear phase incursion,
$\bar{n}_{0}(t)=|\alpha(t)|^{2}$ is the average density of the initial number of
photons, and
$\langle\hat{N}_{{\mathcal{T}}}(t,z)\rangle={\mathcal{T}}|I_{1}(t,z)|^{2}$.

Let us assume that the Green function G(t) has the form
\begin{equation}\label{green}
G(t,z)=(-i2\pi k_{2}z)^{-1/2}\exp{\left(-\frac{it^{2}}{2k_{2}z}\right)},
\end{equation}
where $z$ is the distance traveled in the dispersive linear medium. We have the
coefficients $k_{2}>0$ for normal dispersion of the group velocity, and $k_{2}<0$ for
anomalous dispersion. Expression (\ref{green}) corresponds to the case when
dispersive properties of a medium are taken into account in the second approximation
of the dispersion theory \cite{Vysloukh}.

The dispersion effect is known \cite{Dyakov} to have a spatial analogue in the form
of diffraction spreading of a wave beam. The quantum description of diffraction in
nonlinear optical processes was considered first by S. A. Akhmanov \emph{et al}. in
Ref.\cite{Belinski2} (see also Ref.\cite{Belinski1}), where it was shown that the
mixing of different angular components of a parametrical amplified beam is able to
cause additional nonclassical effects. One can expect similar effects in the case
under consideration. The self-interaction effect of a pulse given by expression
(\ref{oo}) is accompanied by changes in its phase (and therefore frequency) over
time. In the course of its passage through dispersive linear elements, a
frequency-modulated pulse of this type is compressed or stretched \cite{Vysloukh}. It
is precisely this effect that is able to change its photon statistics.

Let us determine the parameter $Q(t,z)$. To simplify calculations, we take the
nonlinearity response function in the form
$h(t)=(1/\tau_{r})\exp{(-t^{2}/\tau^{2}_{r})}$. Let a pulse be initially of the
Gaussian form $\bar{n}_{0}(t)=\bar{n}_{0}\exp{\left(-t^{2}/\tau^{2}_{p}\right)}$. In
the paraxial approximation, i.e., in the case where the phase $\psi(t)$ is replaced
with $\psi_{0}(1-t^{2}/\tau^{2}_{p})$, ($\psi_{0}=2\gamma\bar{n}_{0}$), we obtain for
$\tau_{p}>\tau_{r}$:
\begin{equation}
\langle\hat{N}_{{\mathcal{T}}}(t,z)\rangle=\bar{n}_{0}{\mathcal{T}}V^{-1}(z)
\exp{\left(-\frac{t^{2}}{V^{2}(z)\tau^{2}_{p}}\right)},\label{average}
\end{equation}
\begin{eqnarray}
Q(0,z)=-&\psi_{0}&\left[\frac{{\mathcal{T}}}{\tau_{p}}\right]
\sin\biggl\{\arctan{\left(\frac{\varphi(z)}{\varpi(z)}\right)}\nonumber\\
&+&\frac{1}{2}\arctan{\biggl(\frac{2\varphi_{d}(z)\varpi(z)}{2\varphi(z)\varphi_{d}(z)-\varpi^{2}(z)}}\biggl)\biggl\}\nonumber\\
&\biggl/&\!\left[\varpi^{4}(z)-2\varphi^{2}(z)\varpi^{2}(z)+
4\varphi^{4}(z)\right]^{1/4}\!,\label{last}
\end{eqnarray}
where
\begin{eqnarray}\label{terms}
V^{2}(z)=\varpi^{2}(z)+\varphi^{2}(z); \quad&{}&
\varpi(z)=1-s\psi_{0}\varphi(z);\nonumber\\ \varphi(z)=\frac{z}{D};\quad
\varphi_{d}(z)=\frac{z}{d}; \quad&{}& D=\frac{\tau^{2}_{p}}{|k_{2}|};\quad
d=\frac{\tau^{2}_{r}}{|k_{2}|};
\end{eqnarray}
$D$ and $d$ are the characteristic dispersion lengths, $s=1$ for $k_{2}<0$, and
$s=-1$ for $k_{2}>0$. Let us restrict our consideration to the analysis of the
parameter $Q(t,z)$ for $t=0$, because expression (\ref{last}) for an arbitrary moment
of time $t$ is rather cumbersome.

\begin{figure}[t]
   \vspace{-1cm}
   \begin{center}
       \leavevmode
       \epsfxsize=.48\textwidth
       \epsfysize=.4\textwidth
       \epsffile{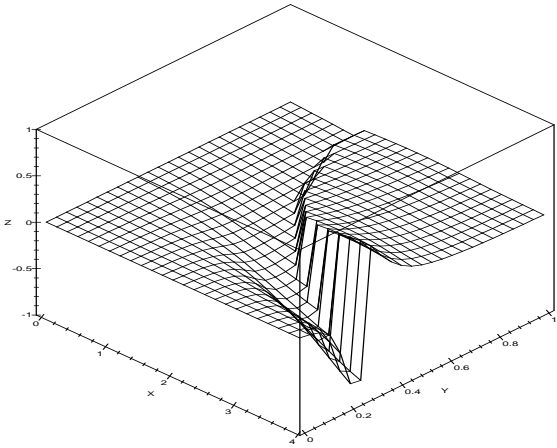}
   \end{center}
   \caption{Dependence of the parameter $Q(0,z)$ [Z] in the case of compression of a USP
with SPM on the nonlinear phase incursion $\psi_{0}$ [X] and the dispersion phase
$\varphi(z)$ [Y] for $\tau_{p}/{\mathcal{T}}=8$.\label{fig1}}
\end{figure}

\begin{figure}[t]
    \vspace{-1cm}
    \begin{center}
        \leavevmode
        \epsfxsize=.48\textwidth
        \epsfysize=.4\textwidth
        \epsffile{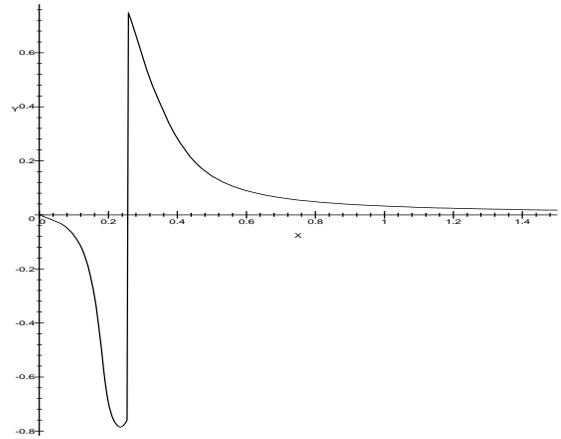}
    \end{center}
\caption{Dependence of the parameter $Q(0,z)$ [Y] in the case of compression of a USP
with SPM on the dispersion phase $\varphi(z)$ [X] for $\psi_{0}=4$ and
$\tau_{p}/{\mathcal{T}}=8$.\label{fig2}}
\end{figure}

\begin{figure}[t]
    \vspace{-1cm}
    \begin{center}
        \leavevmode
        \epsfxsize=.48\textwidth
        \epsfysize=.4\textwidth
        \epsffile{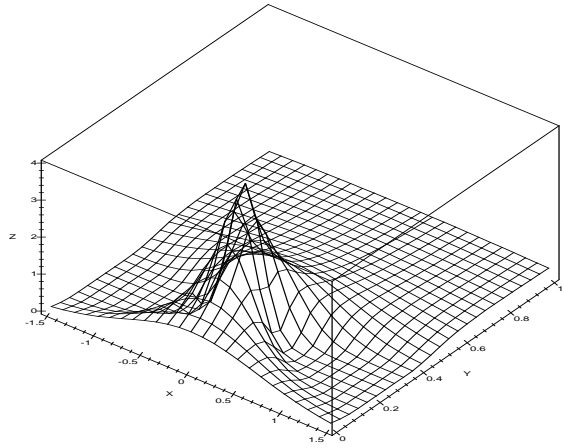}
    \end{center}
\caption{Variation of the shape of a USP with SPM during compression. The calculation
are made in the paraxial approximation.
([Z]$\equiv\langle\bar{N}_{{\mathcal{T}}}(t,z)\rangle$, [X]$\equiv t/\tau_{p}$,
[Y]$\equiv\varphi(z)$). \label{fig3}}
\end{figure}

{}From (\ref{average}) and (\ref{last}) it follows that one can have $Q(t,z)<0$,
i.e., obtain a pulse with sub-Poissonian photon statistics through both compression
($s=1$) and spreading ($s=-1$) of a pulse. Of particular interest from the practical
point of view is the case of anomalous group velocity dispersion ($s=1$) in which the
compression of a phase-modulated pulse (see (\ref{average})) takes place. The
dependence of the Mandel parameter and the average number of photons in this case are
presented in Fig.\ref{fig1}-Fig.\ref{fig2} and Fig.\ref{fig3}.
One can see that the suppression of fluctuation of the number of photons becomes
noticeable for the nonlinear phase $\psi_{0}>1$. For a given nonlinear phase
$\psi_{0}$, there is a certain dispersive phase incursion for which the
suppression of the number of photons is the greatest. According to (\ref{average}),
this is obtained for $\varphi_{opt}(z)=1/\psi_{0}$. In this case, the Mandel
parameter has the minimum value
$Q_{min}=-\left[{\mathcal{T}}/\tau_{p}\right]\psi^{2}_{0}$. Note that a pulse has the
minimal duration for $\varphi(z)=\psi_{0}/(1+\psi^{2}_{0})$. For $\psi_{0}\gg1$ both
extremities are almost coincident.

As follows from calculations, the photon statistics for the measurement time
${\mathcal{T}}\gg\tau_{p}$ is of the Poissonian type. A situation identical in many
aspects to the one considered here was observed in spectral measurements of
fluctuations of femtosecond pulses transmitted through an optical fibre in the region
of normal group-velocity dispersion. An increase in the spectral bandwidth of a
filter caused an increase of the level of fluctuations \cite{Konig}.

Let us present a numerical example. Consider the interaction of a pulse at a
wavelength of $1$ $\mu m$ in an optical fibre, and let its initial duration and
maximum intensity be $2\tau_{p}=10~ps$ and $10^{7}~W\!\cdot\! cm^{-2}$. In this case,
the nonlinear phase in a quartz fibre for the path length $l=100~m$ is $\psi_{0}=3$.
If the resulting USP travels through an optical compressor with anomalous dispersion
$|k_{2}|=10^{-26}~s^{2}\!\cdot cm^{-1}$, the dispersion phase
$\varphi_{opt}(z)=1/\psi_{0}$ is obtained at a distance of about $8~m$. This example
illustrates the feasibility of formation of USPs with sub-Poissonian photon
statistics by the method proposed above.

The obtained result  shows that it is possible in principle to form a pulse
with sub-Poissonian photon statistics in the case where a high-intensity pulse
undergoes SPM during its travel through a nonlinear medium and subsequently passes a
dispersive optical device (an optical compressor or an optical fibre). In view of the
fact that these processes occur in succession, one can choose an optimum scheme or
optimum conditions at each transformation stage. In the specific case of formation of
low-intensity USPs in a nonclassical state, one can attenuate high-intensity pulses
at the output of a nonlinear medium. It is natural that this is accompanied by a
partial loss of nonclassical properties. It should be noted that in the case of
formation of optical solitons, SPM and the dispersion effect take place simultaneously
\cite{Vysloukh}.
\medskip

{\sf Authors are grateful to V. A. Vysloukh for fruitful discussions. This
work was supported in part by the Fundamental Metrology Program of the State
Committee on Science and Technology.}

\end{document}